\documentclass[twocolumn,floatfix,prb,aps,showpacs,superscriptaddress]{revtex4}
\usepackage{epsfig,graphicx}
\usepackage{flafter}
\usepackage{amsmath}
\usepackage{color}
\usepackage{bm}
\usepackage{amssymb}
\usepackage{epstopdf}
\usepackage{amstext}
\usepackage{amsthm}
\usepackage{amsfonts}
\usepackage{latexsym}
\usepackage{multirow}
\usepackage{mathtools}
\usepackage{bm}
\usepackage{bbm}

\begin{document}

\title{Interaction-enhanced flow of a polariton persistent current in a ring}

\author{A. Gallem\'{\i}}
\affiliation{Departament de F\'{\i}sica Qu\`antica i Astrof\'{\i}sica, Universitat de Barcelona, E--08028 Barcelona, Spain}
\affiliation{Institut de Nanoci\`encia i Nanotecnologia de la Universitat de Barcelona, IN$\,^2$UB, E--08028 Barcelona, Spain}
\author{M. Guilleumas}
\affiliation{Departament de F\'{\i}sica Qu\`antica i Astrof\'{\i}sica, Universitat de Barcelona, E--08028 Barcelona, Spain}
\affiliation{Institut de Nanoci\`encia i Nanotecnologia de la Universitat de Barcelona, IN$\,^2$UB, E--08028 Barcelona, Spain}
\author{M. Richard}
\affiliation{Universit\'e Grenoble Alpes, and CNRS,  Institut N{\'e}el, 38000 Grenoble, France}
\author{A. Minguzzi}
\affiliation{Universit\'e Grenoble Alpes, and CNRS, LPMMC, 38000 Grenoble, France}

\date{\today}

\begin{abstract}
We study the quantum hydrodynamical features of exciton-polaritons flowing circularly in a ring-shaped geometry. We consider a resonant-excitation 
scheme in which the spinor polariton fluid is set into motion in both components by spin-to-orbital angular momentum conversion. We show that this 
scheme allows to control the winding number of the fluid, and to create two circulating states  differing by two units of the angular momentum. We 
then consider the effect of  a disorder potential, which is always present in realistic nanostructures. We show that a smooth disorder is efficiently 
screened by the polariton-polariton interactions, yielding a signature of polariton superfluidity. This effect  is reminiscent of supercurrent in a 
superconducting loop.
\end{abstract}


\maketitle

Superfluidity is a striking feature of quantum fluids. It is characterized by an irrotational particle flow, which is frictionless below a critical 
velocity. Superflow is a typical manifestation of a superfluid: when the latter is trapped in a ring and set in circular motion, it will exhibit 
(i) an integer angular momentum in units of $\hbar$ and (ii) a vanishing decay of the current. This phenomenon has been observed a long time ago 
in a superconducting loop below the critical current \cite{File1963}, with superfluid Helium \cite{Reppy1964} and more recently in ultra-cold atom 
condensates \cite{Ramanathan2011}.

Exciton-polaritons, in spite of their nonequilibrium character have also been found to display many features of superfluidity, like frictionless 
flow \cite{Amo2009}, quantized vortices \cite{Lagoudakis2008,Lagoudakis2009,Sanvitto2010}, and Bogoliubov dispersion \cite{Kohnle2011}. A specific 
feature of polaritons is the fact that the superfluid can be excited resonantly both in terms of phase and amplitude. As a result, nontrivial flow 
patterns with finite angular momentum have been imprinted and studied \cite{Boulier2016}. Moreover, polaritons benefit from a spin-orbit coupling 
allowing for spin-to-orbital angular momentum conversion \cite{Manni2011}.

In this work we examine theoretically the polaritonic counterpart of a persistent current in a loop, i.e. in a ring-shaped confining geometry. Polaritonic 
microcavities etched into complicated shapes, such as rings, can be experimentally realized nowadays with a high degree of accuracy using state-of-the-art 
semiconductor nanotechnology \cite{Jacqmin2014,Marsault2015,Kim2011}. In order to excite the circular motion of the polariton fluid, we rely on 
the specific spin-to-orbital angular momentum conversion mechanism. As a result, the angular momentum achieved by the polariton fluid is not directly 
imprinted by the excitation laser phase pattern. We then investigate the competition between this angular momentum generation mechanism, and the 
backscattering due to disorder within the ring.


In presence of disorder, Bose fluids are subject to localization, i.e. Anderson localization for vanishing interaction \cite{Anderson1958}, or many-body 
localization in the strongly interacting case \cite{Basko2006,Basko2007}. These localization mechanisms hinders the quantum fluid flow. However, repulsive 
interactions also screen the disorder experienced by the fluid, which on the contrary, helps restoring the flow, such that the net effect of interactions 
in presence of disorder is in general not easy to determine. Note that such a screening effect has been reported already in a disordered polariton 
condensate \cite{Baas2008} and in ultracold atoms in harmonic traps \cite{Deissler2010} (see e.g. \cite{Modugno2010} for a comprehensive review).

The effect of disorder on persistent currents has been the object of intense studies for fermionic systems, for negligible interactions (see e.g. Ref. 
\cite{Bleszynski2009}, and references therein), as well as including interaction effects \cite{Filippone2016}. In this work, we examine the case of a 
bosonic quantum fluid in driven-dissipative conditions confined within a ring-shaped trap of finite thickness. We show that while the build-up of a net 
polariton flow (i.e. angular momentum) is prevented at large disorder amplitude and weak polariton-polariton interaction, it is restored by increasing 
the interactions.

The paper is organized as follows. In Section \ref{sect:model} we introduce the model that describes the polarization-dependent polariton field, including 
the transversal electric-transversal magnetic (TE-TM) splitting which is present in realistic polaritonic microstructures. Section \ref{sect:soam} describes 
the mechanism of spin-to-orbital angular momentum conversion using numerical simulations of the coupled driven-dissipative Gross-Pitaevskii equations for 
the ring-trapped condensate. In Section \ref{sect:disorder}, the interplay between interactions and disorder is analyzed, and the suppression of 
persistent currents is estimated. Finally, in Section \ref{sect:conclusions} we summarize our results and discuss perspectives.

\section{The model}
\label{sect:model}

Polaritons are bosonic quasi-particles of mixed exciton-photon nature, that exist in semiconductor microcavities in the strong coupling regime \cite{Carusotto2013}. 
%
%
In this work, we consider the lower polariton state, the dispersion of which is well described by a two-coupled harmonic oscillator model as $E_{\rm pol}=\frac{1}{2}
\left(\hbar \omega_c(k)+E_{\rm x}\right)-\frac{1}{2}\sqrt{\left(\hbar \omega_c(k)-E_{\rm x}\right)^2+4\Omega^2}$, where $\Omega$  is  the exciton-photon Rabi splitting, 
$E_{\rm x}$ is the exciton energy, and $\hbar\omega_c(k) \simeq E_{\rm c}+\hbar^2k^2/2m_{\rm eff}$ is the bare cavity photon dispersion characterized by an effective 
mass $m_{\rm eff}$ that typically amounts to $10^{-5}$ in free electron mass units, and $E_{\rm c}$, which is the photonic zero-point kinetic energy.

\begin{figure}[t!]
\centering
\vskip1mm
\includegraphics[width=\linewidth]{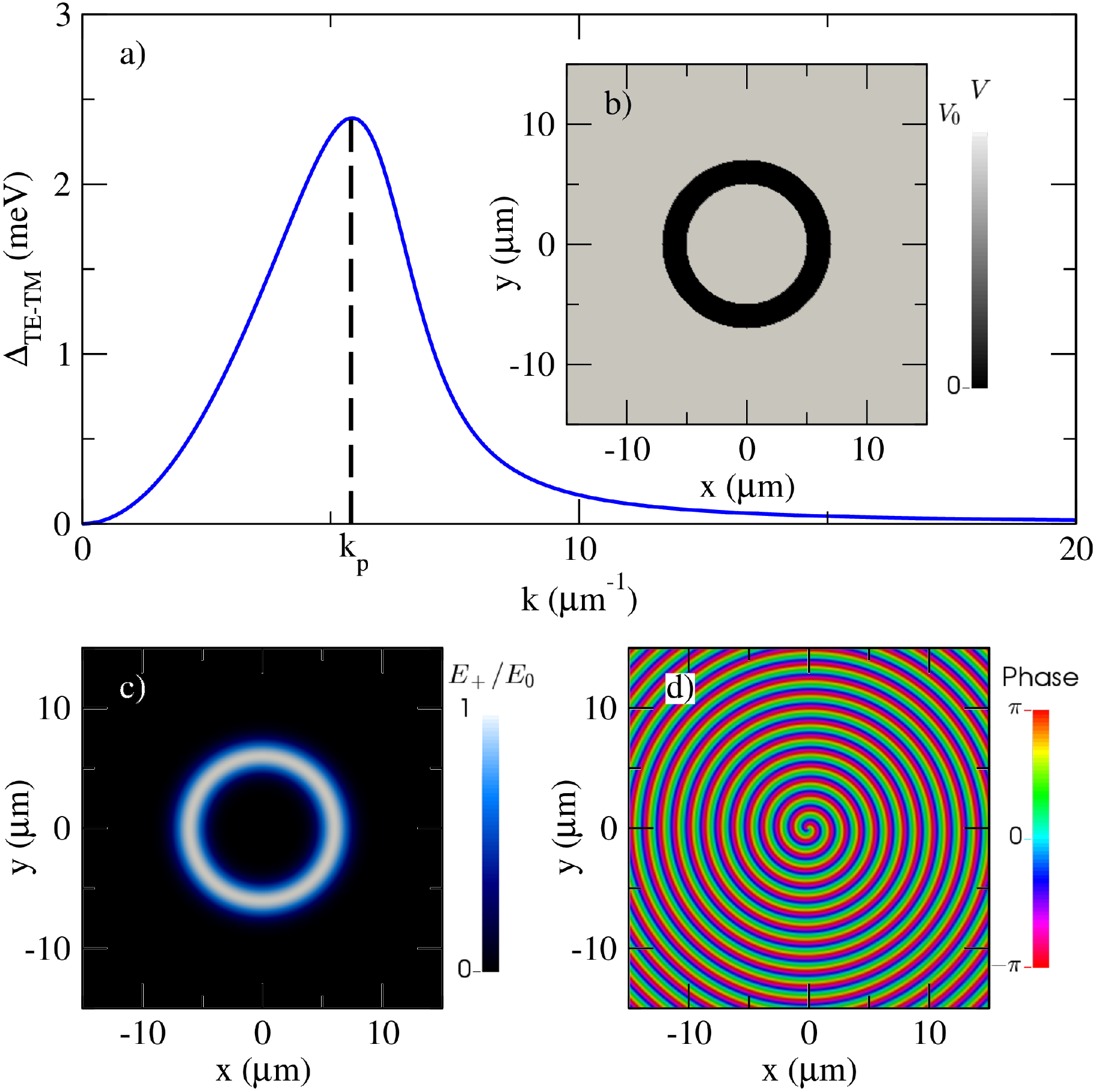}
\caption{
The panel (a) shows the TE-TM splitting as a function of the wave vector and the dashed line points the value of the wave vector of the pump, which is $k_p=5.2\,\mu m^{-1}$. 
Panel (b) depicts the potential of the system used in Sect. \ref{sect:soam}. Panels (c) and (d) show the intensity and phase of the Laguerre-Gauss pump, respectively, for a 
winding number of the pump equal to $1$. }
\label{Fig1}
\end{figure}

In the following, we include the cavity photon polarization degree of freedom in our description in terms of a pseudo-spin by means of the components of the Stokes vector. 
Our discussion will involve two polarization basis: the circular polarization basis $|\pm\rangle$ relevant to polariton-polariton interactions, and the horizontal-vertical 
linear polarization basis $ |h\rangle$, $|v\rangle$ which is important in order to account for the TE-TM splitting of the cavity mode. The two basis are related by 
a rotation according to the usual transformation $|\pm\rangle =(|h\rangle\pm i  |v \rangle)/\sqrt{2}$.

Indeed, owing to the Fresnel relation, TE and TM polarized light experience a slightly different optical path within the cavity, which gives rise to a slightly different 
effective mass $m_{\rm TE}$, $m_{\rm TM}$ for both polarization states. Figure \ref{Fig1}(a) shows the TE-TM splitting \cite{Panzarini1999} versus wave vector $k$, with 
the following parameters: $m_{\rm TE}=1.94\times10^{-5}m_{\rm e}$, $m_{\rm TM}=2.06\times10^{-5}m_{\rm e}$, $E_{\rm c}=2750$ meV, $E_{\rm x}=2820$ meV and $\Omega=30$ 
meV. These parameters have been chosen as to match ZnSe-based microcavities with which we plan to implement an experimental realization of this proposal.

In the simulations that we will present in the following sections, the radial momenta are discretized due to the confinement within the ring. The TE and TM modes, having 
a different effective mass, are thus split. We maximize the effect of the TE-TM splitting by applying a radial momentum $k_p=5.2\,\mu m^{-1}$ to the pump beam. 

We now define $\psi_\alpha(\vec r,t)$ as the polariton field with a polarization state $\alpha$. The corresponding dynamics is determined by two coupled driven-dissipative 
Gross-Pitaevskii equations \cite{Kavokin2004}. In the circular polarization basis it reads:
\begin{widetext}
\begin{equation}
i\hbar \frac{\partial}{\partial t}\psi_\alpha(\vec r, t)=[T^{\rm circ}_{\alpha\alpha}(\vec k\,)+V(\vec r\,)-i\hbar\gamma_\alpha] \psi_\alpha(\vec r, t)+T^{\rm circ}_{\alpha\beta}(\vec k\,)\psi_\beta(\vec r,t)+
(g_{\alpha\alpha}|\psi_\alpha(\vec r, t)|^2+g_{\alpha\beta}|\psi_\beta(\vec r, t)|^2)\psi_\alpha(\vec r, t)+i E^{\rm field}_{\alpha}(\vec r, t)\,,
\label{eq2}
\end{equation}
\end{widetext}
where $T^{\rm circ}_{\alpha\beta}(\vec k\,)$ is the kinetic tensor in the circular polarization basis, $V(\vec r\,)$ is an external potential, which includes the ring 
confinement and an optional disorder potential, and $g_{\alpha\alpha}$ ($g_{\alpha\beta}$) is the intercomponent (intracomponent) interaction strength. The subindices 
\mbox{$\alpha,\beta={+,-}$}; $\alpha\neq\beta$ describe the different polarization components of the polariton field. The driven-dissipative features are explicitly included 
by means of the loss rate $\gamma_\alpha=1/\tau_\alpha$ describing polaritons leaking throughout the microcavity mirrors, where $\tau_\alpha=2\,\mbox{ps}$ is the polariton 
lifetime, and a coherent pump term $E^{\rm field}_{\alpha}(\vec r, t)$ that injects polaritons resonantly.

For the kinetic tensor, we use \cite{Kavokin2004}
\begin{equation}
T^{\rm circ}(\vec k\,)=\begin{pmatrix}
\hbar \omega & & \Delta\,\displaystyle\frac{(-k_x+i \,k_y)^2}{k^2} \\
 \Delta\,\displaystyle\frac{(-k_x-i \,k_y)^2}{k^2}& & \hbar \omega
\end{pmatrix}
,
\label{eq4}
\end{equation}
where $\hbar\omega=(\hbar\omega_{\rm TM}+\hbar\omega_{\rm TE})/2$ is the average energy between the TE and the TM cavity modes, and \mbox{$\Delta=\hbar (\omega_{\rm TM}-\omega_{\rm TE})/2$} 
is the TE-TM splitting. Notice that the off-diagonal terms effectively play the role of a spin-orbit coupling.

\section{Spin-to-orbital angular momentum conversion}
\label{sect:soam}

One of the most striking effects that arise from the TE-TM splitting is the possibility to generate vortices by effective spin-orbit 
coupling, which leads to a spin-to-orbital angular momentum (SOAM) conversion. It means that we can excite a polaritonic field of a 
given polarization components with zero-angular momentum, and obtain a vortex with winding number two in the cross-polarized component. 
This effect was \mbox{theoretically} predicted in Ref. \cite{Liew2007}, and experimentally confirmed in Ref. \cite{Manni2011} in a homogeneous 
two-dimensional semiconductor. We present an alternative derivation of this effect in \mbox{Appendix~\ref{appA}}. In this section, we analyze 
this effect in a ring-shaped trap for polaritons. In this case, the vortex appears as a persistent current along the ring which is 
reminiscent of a persistent current of a superfluid within a loop.

\subsection{Spin-to-orbital angular momentum conversion in ring-shaped traps}

We use the following potential to describe the ring-shaped trap in the driven-dissipative Gross-Pitaevskii equation (\ref{eq2}):
\begin{equation}
V(r)=V_0\left(1-\frac{\sinh(w/\xi)}{\cosh(w/\xi)+\cosh((r-R_0)/\xi)}\right)\,,
\end{equation}
where $r$ is the two-dimensional radial coordinate. This potential corresponds to a ring with mean radius \mbox{$R_0=6\mu\mbox{m}$}, width $w=1\mu\mbox{m}$ and depth $V_0=1$ eV. 
The profile of the edges of the trap is described by the parameter $\xi$, which we fix to be $w/10$. The potential is represented in Fig. \ref{Fig1}(b).

\begin{figure}[t!]
\centering
\vskip1mm
\includegraphics[width=\linewidth]{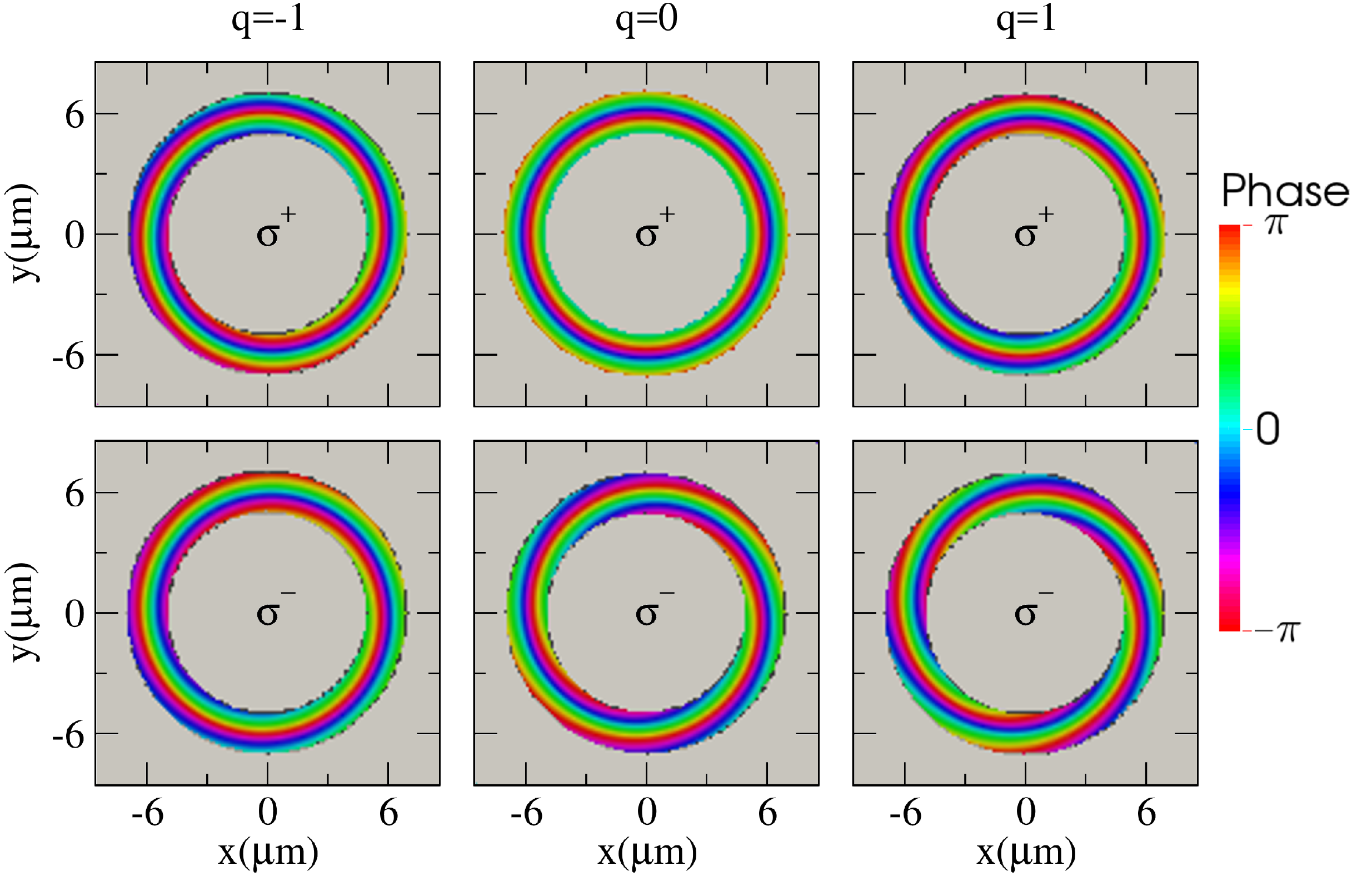}
\caption{Phase profile of the $\sigma^+$ (top row) and the $\sigma^-$ (bottom row) components, in the steady state for different values of the winding number associated to the 
Laguerre-Gauss beam of the pump. In the left panels, $q=-1$, in the middle panels $q=0$ and in the right panels $q=1$. The winding number of the persistent current nucleated in 
the $\sigma^-$ component is $2$ units larger than that of the pump.}
\label{Fig2}
\end{figure}

The pump geometry is illustrated in the panels (c) for the intensity and (d) for the phase of Fig. \ref{Fig1}: it consists in a ring-shaped Laguerre-Gauss mode with a radial 
phase dependence $k_pr$ plus an azimuthal one $q\theta$, where ${\vec r}=(r,\theta)$ that reads
\begin{equation}
E_+^{\rm field}(\vec r\,)=E_0\, e^{-\frac{(r-R_0)^2}{2w^2}}e^{ik_pr}e^{iq\theta}\,,
\end{equation}
where $E_0=10$ meV/$\mu$m is the amplitude of the pump, which pumps polaritons with  $\sigma^+$ polarization directly within the ring. In addition, it can imprint orbital angular momentum to 
polaritons, in the same way as in Ref. \cite{Sanvitto2010}, where the authors used this property to generate a vortex.

We have numerically solved this two-dimensional driven-dissipative Gross-Pitaevskii equation to obtain the steady state of the system. In Fig. \ref{Fig2} we plot the phase profile 
of the polariton field of the $\sigma^+$ (top row) and the $\sigma^-$ (bottom row) component inside the ring (i.e. for $|r-R_0|<w$, where the density does not vanish). We represent 
the case where the winding number of the angular momentum carried by the pump is $q=-1$ in the left column, $q=0$ in the middle column, and $q=1$ in the right column. As expected, 
in the stationary state, the component co-polarized with the pump exhibits a persistent current with a phase winding number \mbox{matching} the pump. Interestingly, we find that 
the cross-polarized component exhibits a persistent current with winding number $1$, $2$ and $3$ for the left, middle and right \mbox{column}, respectively. This is a two units 
increase with respect to the winding number of the pump. We show in \mbox{Appendix~\ref{appA}} that the same feature actually occurs also in the homogeneous  system. It is also interesting to notice that 
the large radial component of the phase gradient in each ring results from the spin-orbit coupling between the two spin components. Correspondingly, we find a modulation in the radial density profile of each component, which is due to the coupling between the transverse modes.

\begin{figure}[t!]
\centering
\vskip1mm
\includegraphics[width=\linewidth]{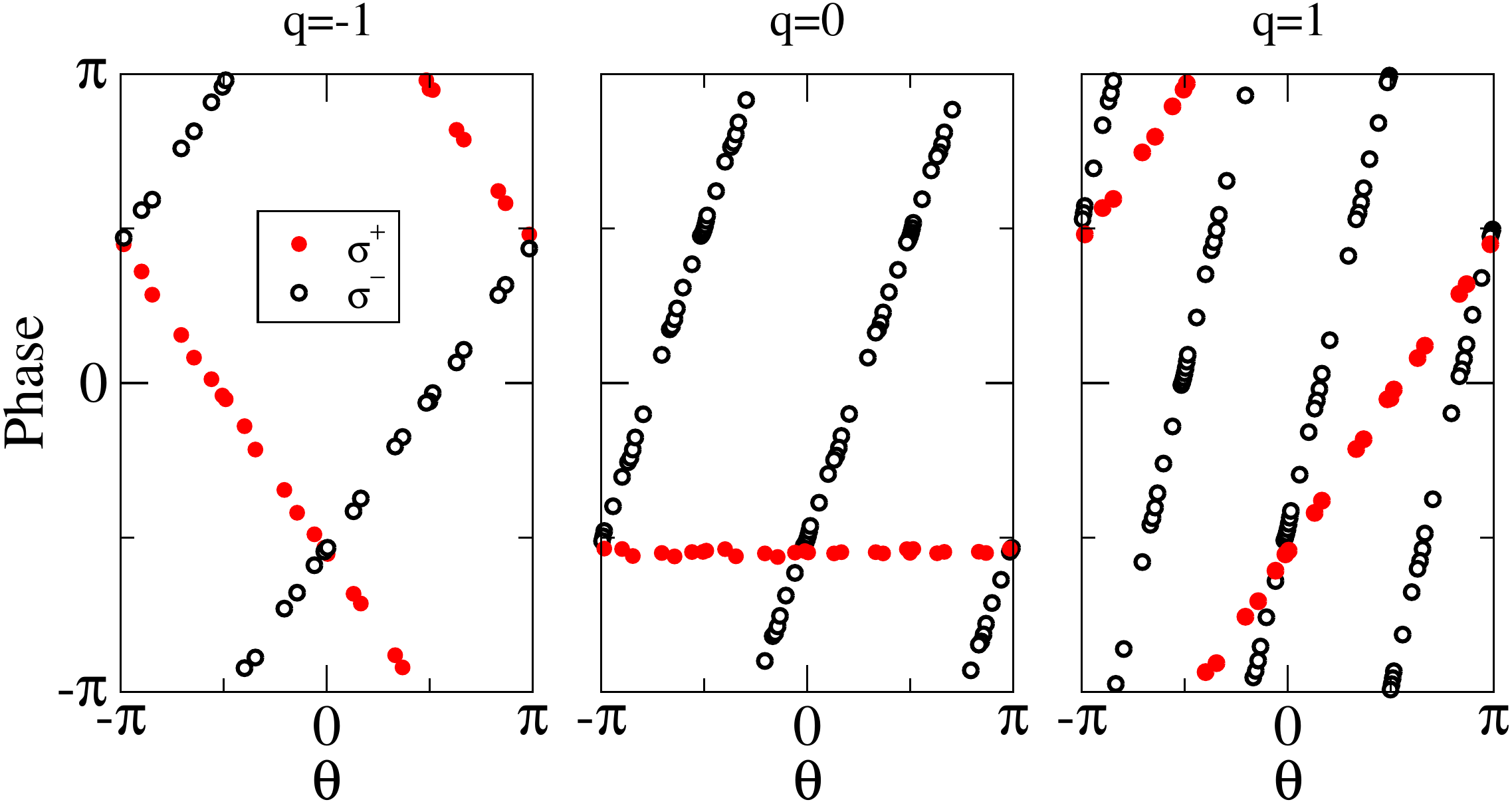}
\caption{Plot of the phase of the $\sigma^+$ (red filled circles) and $\sigma^-$ (black open circles) components as a function of the angle $\theta$ along the ring, at $r=R_0$ 
for a Laguerre-Gauss pump with orbital angular momentum with winding number $-1$ (left panel), $0$ (middle panel) and $1$ (right panel).}
\label{Fig3}
\end{figure}

The spin-to-orbital angular momentum conversion
can be also seen in Fig. \ref{Fig3}, where we show the phase $\phi(r=R_0,\theta)$ of the $\sigma^+$ polarized field (red filled circles) and of the $\sigma^-$ (black open circles) 
one in the steady state. The winding number of the pump is $q=-1$ (left panel), $q=0$ (middle panel) and $q=1$ (right panel). We see that the phase of the $\sigma^-$ component winds 
by $2\pi$, $4\pi$ and $6\pi$, respectively, i.e. $q^-=1,2$ and $3$ as expected from Fig. \ref{Fig2}.

%


\section{Polariton Current: Competition between disorder and interactions}
\label{sect:disorder}

\begin{figure}[t!]
\centering
\vskip1mm
\includegraphics[width=0.9\linewidth]{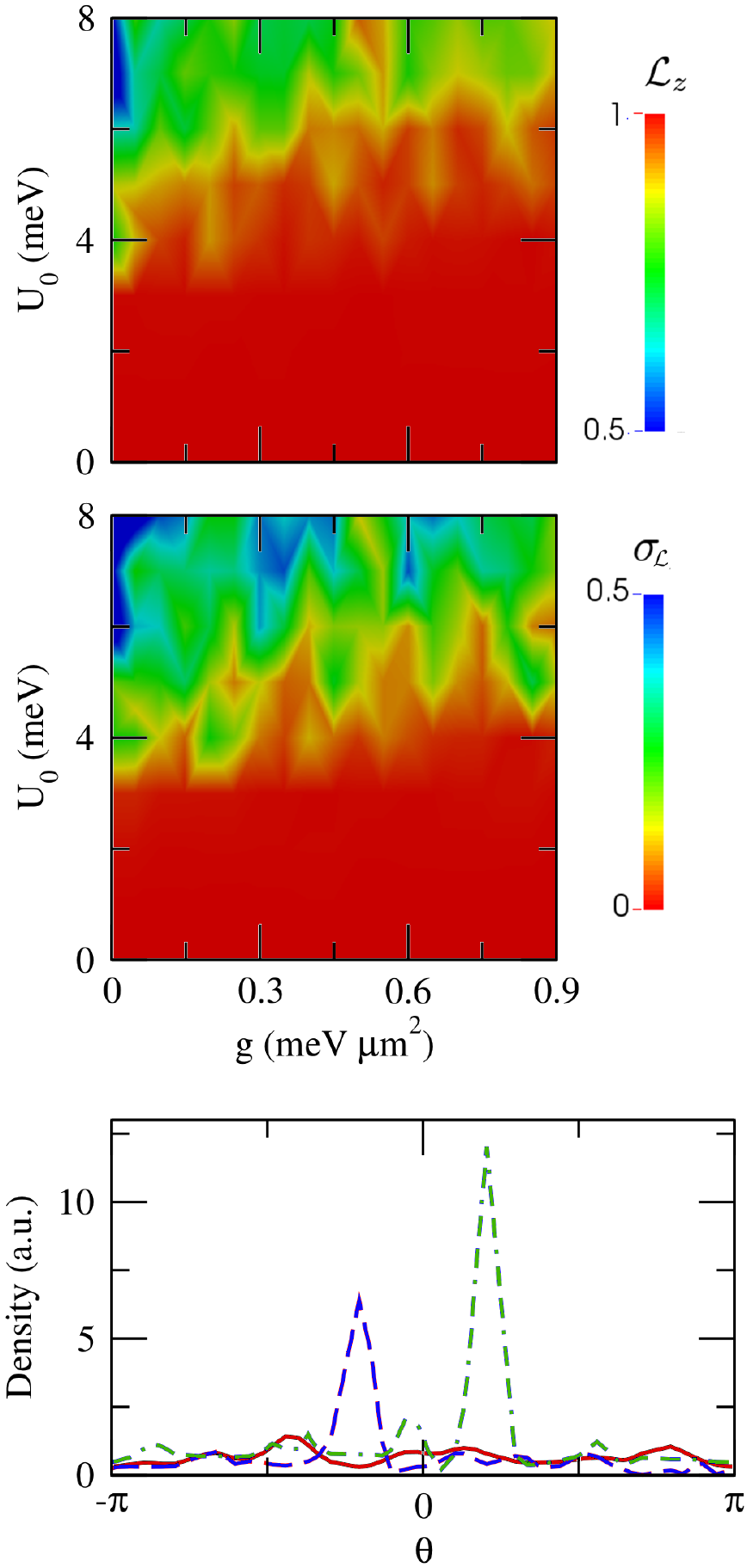}
\caption{Top panel: magnitude of the angular momentum  ${\cal L}_z$ (normalized to its maximum value) as a function of the interaction $g$ and the disorder strength $U_0$. The value represented in the color map is the 
result of the average of ten dynamical simulations of the driven-dissipative Gross-Pitaevskii equation. Middle panel: standard deviation of ${\cal L}_z$ in the  numerical calculations. 
The bottom panel represents the density profile (in arbitrary units) along the ring at $r=R_0$ for different values of the interaction constant $g$ and disorder strength $U_0$: $g=0$ 
meV$\mu$m$^2$ and $U_0=5$ meV (dot-dashed green line), $g=0.6$ meV$\mu$m$^2$ and $U_0=5$ meV (solid red line), and $g=0.6$ meV$\mu$m$^2$ and $U_0=8$ meV (dashed blue line).}
\label{Fig4}
\end{figure}


In the previous section, we have shown that when we excite the $\sigma^+$ polariton component into a mode carrying no angular momentum in a smooth ring-shaped trap, the $\sigma^-$ component 
acquires a persistent current with winding number $2$. In this section we account for the fact that in realistic experiments, a (gaussian-distributed) disorder potential 
\begin{equation}
V_{\rm dis}(\vec r\,)=\mbox{Re}\left[\mathcal{F}^{-1}[2\pi^2 l_c U_0 \exp(i\varphi_{\vec k})\exp(-k^2 l_c^2/4)](\vec r\,)\right]\,,
\end{equation}
experienced by polaritons is present within the ring, where $U_0$ is the strength of the disorder, $l_c$ is the correlation length, which gives the order of magnitude of the distance between 
maxima and minima of the disorder potential, and $\varphi_{\vec k}$ is a random matrix with phases uniformly distributed between $0$ and $2\pi$. We also analyze the interplay of disorder and 
interactions on the polariton current along the ring.

In the simulations, in agreement with the literature, we have fixed the polariton-polariton interaction to be 10 times larger in the co-polarized 
case than in the cross-polarized, i.e. $g=g_{++}=g_{--}=10\,g_{+-}$. We use the pump to excite the $\sigma^+$ component with a Laguerre-Gauss beam 
with $q=0$ orbital angular momentum. To monitor the persistent current induced in the $\sigma^-$ component, we compute the expectation value of the 
angular momentum operator around the $z$-axis ${\cal L}_z=\langle \hat{L}_z\rangle$, normalized to its maximum value.


In the top panel of Fig. \ref{Fig4}, we show  the angular momentum computed from the average of ten dynamical simulations performed with different 
realizations of the disorder potential, as a function of the interaction strength $g$ and disorder strength $U_0$. We can see from the figure that 
there is a (red) region in which the disorder can be simply ignored, since it does not affect the polariton field, the current is preserved and the 
system remains superfluid. However, as the disorder strength increases above a given critical value (yellow region), the polariton persistent current 
diminishes. Moreover, larger interactions require larger values of the disorder strength in order to observe the decrease of such a current. The 
reason of this effect is the fact that the disorder is efficiently screened by the interactions. On the contrary, the polariton current is suppressed 
as disorder overcomes interaction. 

For each point of the top panel, we have computed the standard deviation $\sigma_{\mathcal{L}}$ of the different values of ${\cal L}_z$ obtained for 
each realization of the disorder. The result is represented in the middle panel of Fig. \ref{Fig4}. We see that when a persistent current exists, 
${\cal L}_z$ does not fluctuate much from one realization to the next. Whereas in the regime where the disorder and the interactions compete equally, 
the actual ${\cal L}_z$ which is achieved is highly dependent on the details of $V_{\rm dis}(\vec r\,)$, and the standard deviation increases.


The simulations show also that in the non-interacting regime, density hot spots build up as a result of Anderson localization, and the current along 
the ring is thus substantially reduced. An example of this regime is shown in the bottom panel of Fig. \ref{Fig4} where the density profile at $r=R_0$ 
is shown for $g=0$ and $U_0=5$ meV (dot-dashed green line). Then, upon increasing the interactions to $g=0.6$ meV$\mu$m$^2$, the flow is restored and 
the polariton density becomes much more homogeneous within the ring (solid red line). This flow can be suppressed again by increasing the disorder 
amplitude to $U_0=8$ meV. In this case, the polariton density exhibits another hot spot (dashed blue line).

\section{Conclusions}
\label{sect:conclusions}

In conclusion,  in this work we have demonstrated the generation of persistent currents with arbitrary winding number in a two-component polariton 
condensate by the mechanism of spin-to-orbital angular momentum conversion. This allows to generate with a single pump two persistent current states, 
differing by two units of winding number. Furthermore, we have studied the effect of a possible disorder on the persistent currents. We have identified 
two main regimes at weak interactions, one where the polariton condensate is superfluid and screens the effect of disorder, and one where localization 
effects overcome superfluidity and strongly reduce the persistent currents. The latter regime is expected to occur for very large values of disorder 
strength or weak interactions. This allows us to conclude that one can expect robust persistent current states under typical experimental conditions.

In outlook, it would be interesting to explore the interplay of superfluidity and interactions for bosons in driven-dissipative conditions by going 
beyond the mean-field approximation. Also, it would be interesting to manipulate the two-current condensate for the study of the dynamics of superfluidity.

\section*{Acknowledgments}
We acknowledge financial support from the Spanish MINECO (FIS2014-52285-C2-1-P) and the European Regional development Fund, Generalitat de Catalunya 
Grant No. SGR2014-401. AG is supported by the Spanish MECD fellowship FPU13/02106. AM ackowledges funding from the ANR SuperRing (ANR-15-CE30-0012-02), 
MR acknowledges funding from the ANR QFL (ANR-16-CE30-0021-04).

\appendix

\begin{widetext}

\section{SOAM conversion in homogeneous two-dimensional polariton gas}
\label{appA}

In this section we provide an alternative derivation of the SOAM conversion developed in Ref. \cite{Liew2007} for the case of non-interacting polaritons 
in a homogeneous trap. We also suppose that the lifetime of the polaritons of both polarization components is equal. This is a realistic assumption, since 
the lifetime does not strongly depend on the polarization of the polariton condensate. Under this condition, we can write the coupled driven-dissipative 
Gross-Pitaevskii equations (\ref{eq2}) in $\vec k$-space:
\begin{equation}
i\hbar\frac{\partial}{\partial t}{\Psi}(\vec k ,t)=\left[T(\vec k)-i\hbar\gamma\mathbbm{1}\right]\Psi(\vec k, t)+iE(\vec k, t)\,,
\end{equation}
where $\Psi(\vec k, t)$ and $E(\vec k, t)$ are spinors that contain the polariton field of each polarization component of the polariton condensate, and 
the pump in each component, respectively. The kinetic tensor (\ref{eq4}) can be written as:
\begin{equation}
T(\vec k)=\hbar
\begin{pmatrix}
\omega(\vec k\,) &  \Delta(\vec k\,) e^{-i\,2\phi} \\
\Delta(\vec k\,) e^{i\,2\phi} & \omega(\vec k\,) 
\end{pmatrix}
\,,
\end{equation}
where $k_x=k\,\cos\phi$ and $k_y=k\,\sin\phi$, and we have defined $2\omega(\vec k\,)=\omega_{\rm TM}(\vec k\,)+\omega_{\rm TE}(\vec k\,)$ and 
$2\Delta(\vec k\,)=\omega_{\rm TM}(\vec k\,)-\omega_{\rm TE}(\vec k\,)$. We can write $T(\vec k\,)-i\hbar\gamma\mathbbm{1}$ in its diagonal form 
as $MD(\vec k\,)M^{-1}$, where $D (\vec k\,)$ is the diagonal matrix whose elements are the eigenvalues of $T (\vec k\,)-i\hbar\gamma\mathbbm{1}$: 
$\hbar(\omega (\vec k\,)\pm\Delta (\vec k\,)-i\gamma)$, and $M$ is the change of basis matrix. 

At this point, we can rewrite the spinor field and the pump as $\Psi(\vec k,t)=M\Phi(\vec k,t)$ and $E(\vec k,t)=MG(\vec k,t)$. Since $M$ is the 
matrix that diagonalizes the kinetic tensor, this transformation allows us to decouple the previous system of linear equations:
\begin{equation}
i\hbar\frac{\partial}{\partial t}{\Phi}(\vec k ,t)=D(\vec k)\Phi(\vec k, t)+iG(\vec k, t)\,.
\end{equation}
The solution of the homogenous part is 
\begin{equation}
\Phi_{H}(\vec k,t)=\Phi_0(\vec k)\,\exp(-iD(\vec k\,)t/\hbar)\,,
\end{equation}
where $\Phi_0(\vec k\,)$ is an initial condition for the polariton field, and the solution of the inhomogeneous part is: 
\begin{equation}
\Phi_I(\vec k, t)=\Phi_H(\vec k,t)\int_{0}^t\frac{G(\vec k, t')}{\hbar\Phi_H(\vec k, t')}dt'\,,
\end{equation}
with $\Phi(\vec k, t)=\Phi_H(\vec k, t)+\Phi_I(\vec k, t)$. The solution for $\Psi(\vec k, t)$ is then: 
\begin{align}
\Psi(\vec k, t)=&M\Phi_H(\vec k, t)+M\Phi_I(\vec k, t)=M\Phi_0(\vec k\,)\,\exp(-iD(\vec k\,)t/\hbar)\nonumber\\
&+M\exp(-iD(\vec k\,)t/\hbar)\Phi_0(\vec k\,)\int_0^t\Phi_0(\vec k\,)^{-1}\exp(iD(\vec k\,)t'/\hbar)M^{-1}E(\vec k,t)\, dt'\,.
\end{align}

Due to the presence of the dissipative terms, which remain in the diagonal part of $D(\vec k\,)$, one can see that the first term vanishes 
at long times. The second term of the sum simplifies as: 
\begin{equation}
\Psi(\vec k,t)=\int_0^tM\exp(-iD(\vec k\,)(t-t')/\hbar)M^{-1}E(\vec k,t) dt'=\int_0^t U(\vec k,t-t')\exp(-\gamma t)E(\vec k, t')dt'\,,
\end{equation}
where $U(\vec k,t)=e^{-iT(\vec k\,)t}$ is the time evolution operator corresponding to the kinetic tensor, which can be shown to be:
\begin{equation}
U(\vec k,t)=e^{i\omega(\vec k\,) t}
\begin{pmatrix}
\cos(\Delta (\vec k\,)\, t)&  i\exp(-i2\phi)\sin(\Delta (\vec k\,)\, t) \\
i\exp(i2\phi)\sin(\Delta (\vec k\,)\, t) & \cos(\Delta (\vec k\,)\, t)
\end{pmatrix}
\,.
\end{equation}
When the system is pumping only one of the components of the $\sigma^+$-$\sigma^-$ basis, the spinor corresponding to the pump will be 
$E(\vec k, t)=f(\vec k, t)(1 \,\,,\,\,0)^T$. With the aim of demonstrating the spin-to-orbital angular momentum effect, we will restrict 
the pump to the following shape: $f(\vec k, t)=f_0\delta(k-k_p)\delta(t)$. The polariton field in $\vec k$ can be calculated as: 
\begin{equation}
\begin{pmatrix}
\Psi_+(\vec k, t) \\
\Psi_-(\vec k, t)
\end{pmatrix}
= f_0\exp((i\omega (\vec k\,)-\gamma)t)\delta(k-k_p)
\begin{pmatrix}
\cos(\Delta (\vec k\,) t) \\
i\exp(i2\phi)\sin(\Delta (\vec k\,) t) 
\end{pmatrix}
\,.
\end{equation}
The corresponding polariton field in the real space is the inverse Fourier transform. 
\begin{align}
\begin{pmatrix}
\Psi_+(\vec r, t) \\
\Psi_-(\vec r, t)
\end{pmatrix}
= f_0\exp(-i\gamma t)&\int_0^{2\pi} d\phi\int_0^\infty kdk\,\delta(k-k_p)\exp(i\omega (\vec k\,)\,t)\exp(i\vec k \cdot \vec r\,)
\begin{pmatrix}
\cos(\Delta (\vec k\,) t) \\
i\exp(i2\phi)\sin(\Delta (\vec k\,) t) 
\end{pmatrix}
\nonumber\\
=f_0\,k_p\exp((i\omega (k_p)-\gamma)t)\int_0^{2\pi} &d\phi \exp(i k_p  r \cos(\theta-\phi))
\begin{pmatrix}
\cos(\Delta (k_p) t) \\
i\exp(i2\phi)\sin(\Delta (k_p) t) 
\end{pmatrix}
\,,
\end{align}
where we have used that $\vec k \cdot \vec r= k r \cos(\theta-\phi)$, being $\theta$ and $\phi$ the orientation angles of $\vec r$ and $\vec k$, 
respectively. In order to solve the two integrals (one for each component), the following property of Bessel functions will be useful:
\begin{equation}
\mathcal{J}_n(\zeta)=\frac{1}{2\pi}\int_{-\pi}^{\pi}\exp(i(n\tau+\zeta \sin \tau))d\tau\,.
\end{equation}
The final solution is then: 
\begin{equation}
\begin{pmatrix}
\Psi_+(\vec r, t) \\
\Psi_-(\vec r, t)
\end{pmatrix}
=2\pi f_0\,k_p\exp((i\omega (k_p)-\gamma)t)
\begin{pmatrix}
\mathcal{J}_0(k_p\, r)\cos(\Delta (k_p) t) \\
-i\mathcal{J}_2(k_p\, r)\exp(i2\theta)\sin(\Delta (k_p) t) 
\end{pmatrix}
\,.
\end{equation}

We can see from the previous equation that when we pump one of the components, the polariton field of the other component acquires a phase pattern 
with a winding number $2$, which is a doubly-quantized vortex. This phenomenon has been already theoretically predicted in Ref. \cite{Liew2007}, and 
experimentally observed in Ref. \cite{Manni2011}, in the case of non-trapped polariton condensates.

It is worth to comment the case where instead of pumping at a given modulus of $\vec k$ for all the possible angles in momentum space $\phi$, the 
orientation of $\vec k$ is also fixed. In this case, a term $\delta(\phi-\phi_0)$, where $\phi_0$ is the orientation direction of the pump wave 
vector, should be added to the pump. Then, the integral on $\phi$ when doing the inverse Fourier transform becomes trivial, and the solution is: 
\begin{equation}
\begin{pmatrix}
\Psi_+(\vec r, t) \\
\Psi_-(\vec r, t)
\end{pmatrix}
=2\pi f_0\,k_p\exp((i\omega (k_p)-\gamma)t)\,\exp(ik_p\, r \cos(\theta-\phi_0))
\begin{pmatrix}
\cos(\Delta  (k_p) t) \\
-i\exp(i2\phi_0)\sin(\Delta (k_p) t) 
\end{pmatrix}
\,.
\end{equation}
The previous solution does not content any vortex profile, hence, in order to nucleate a vortex, it is crucial not to fix the wave vector orientation 
and excite all the possible angles in momentum space. As an example, if we pump with the pump wave vector oriented along the $x$-direction, the phase 
pattern of the minority component will be the one of a plane wave travelling in the $x$-direction. 
\end{widetext}

%
%
%
%
%
%

\bibliography{bibtex.bib}

\end{document}